\makeatletter
\@namedef{ver@picins.sty}{9999/99/99}
\makeatother

\documentclass[fleqn]{2023SCGE}
\usepackage{hyperref}
\usepackage{epstopdf}
\DeclareUnicodeCharacter{2212}{\ensuremath{-}}
\usepackage{multirow, makecell}

\usepackage[toc]{multitoc}

\begin{document}

\ensubject{subject}

\ArticleType{Article}
\SpecialTopic{SPECIAL TOPIC: AI for Mechanics}

\title{Implicit factorized transformer approach to fast prediction of turbulent channel flows}{Implicit factorized transformer approach to fast prediction of turbulent channel flows}

\author[1,2]{Huiyu Yang}{}%
\author[1,2]{Yunpeng Wang}{}
\author[1,2]{Jianchun Wang}{wangjc@sustech.edu.cn}

\AuthorMark{Yang H Y}

\AuthorCitation{H. Yang, Y. Wang, and J. Wang}

\address[1]{Department of Mechanics and Aerospace Engineering, Southern University of Science and Technology, Shenzhen 518055, China}

\address[2]{Guangdong Provincial Key Laboratory of Turbulence Research and Applications, \\Southern University of Science and Technology, Shenzhen 518055, China}


\abstract{Transformer neural operators have recently become an effective approach for surrogate modeling of systems governed by partial differential equations (PDEs). In this paper, we introduce a modified implicit factorized transformer (IFactFormer-m) model which replaces the original chained factorized attention with parallel factorized attention. The IFactFormer-m model successfully performs long-term predictions for turbulent channel flow, whereas the original IFactFormer (IFactFormer-o), Fourier neural operator (FNO), and implicit Fourier neural operator (IFNO) exhibit a poor performance. Turbulent channel flows are simulated by direct numerical simulation using fine grids at friction Reynolds numbers $\text{Re}_{\tau}\approx 180,395,590$, and filtered to coarse grids for training neural operator. The neural operator takes the current flow field as input and predicts the flow field at the next time step, and long-term prediction is achieved in the posterior through an autoregressive approach. The results show that IFactFormer-m, compared to other neural operators and the traditional large eddy simulation (LES) methods including dynamic Smagorinsky model (DSM) and the wall-adapted local eddy-viscosity (WALE) model, reduces prediction errors in the short term, and achieves stable and accurate long-term prediction of various statistical properties and flow structures, including the energy spectrum, mean streamwise velocity, root mean square (rms) values of fluctuating velocities, Reynolds shear stress, and spatial structures of instantaneous velocity. Moreover, the trained IFactFormer-m is much faster than traditional LES methods. By analyzing the attention kernels, we elucidate the reasons why IFactFormer-m converges faster and achieves a stable and accurate long-term prediction compared to IFactFormer-o. Code and data are available at: \url{https://github.com/huiyu-2002/IFactFormer-m}.}



\maketitle



\begin{multicols}{2}
\section{Introduction}
Turbulence simulation is an important research area with significant applications in aerospace, energy, and many engineering fields \cite{1,1a,1b,1c}. Common methods for turbulence simulation include direct numerical simulation (DNS), large eddy simulation (LES), and Reynolds-averaged Navier-Stokes (RANS) method. DNS directly solves the Navier-Stokes equations, offering the high accuracy, but it is difficult to use for complex problems with high Reynolds numbers \cite{2,2a}. LES resolves the large-scale flow structures and models the effects of small-scale ones using a subgrid-scale (SGS) model \cite{3,3a}, achieving a balance between accuracy and efficiency. RANS method merely solves the averaged flow field with the modeling of the effects of fluctuating flow field, and is widely used for practical engineering problems \cite{4,4a}. These traditional turbulence simulation methods generally have high computational costs, even the less expensive RANS method, which greatly limits their application.

The introduction of machine learning (ML) techniques is expected to solve this problem \cite{5}. Neural operator (NO) models are considered as an effective method for simulating physical systems governed by partial differential equations (PDEs), due to their theoretical foundation \cite{6}. The trained NO model can make efficient and fast predictions, serving as a lightweight surrogate model. As a pioneer of neural operators, Lu et al. \cite{7} introduced deep operator network (DeepONet), which for the first time employed neural networks to learn operators. Li et al. \cite{8} proposed the Fourier Neural Operator (FNO), which leverages discrete Fourier transforms to perform feature fusion in the frequency domain, significantly enhancing both the model's speed and accuracy. Subsequent works have made a series of improvements based on the two models mentioned above \cite{9,10,11,12,12a,12b,12c,12d,12e}.

In recent years, transformer neural operators have been gradually developed \cite{13,14,15,16,17,17b,17c}. Li et al. \cite{13} applied an attention-based encoder-decoder structure to predict tasks related to systems governed by partial differential equations. Hao et al. \cite{14} proposed a general neural operator transformer that encodes information including initial conditions, boundary conditions, and equation coefficients into the neural network, while extracting the relevant physical fields and their correlations. The transformer neural operators developed in the aforementioned works suffer from high computational costs and excessive memory usage. As a result, a series of subsequent studies focused on investigating the performance of lightweight transformer neural operators. Li et al. \cite{15} proposed a low-rank transformer operator for uniform grids, which significantly reduces computational cost and memory usage by employing axial decomposition techniques to sequentially update features in each direction. Wu et al. \cite{16} proposed an efficient transformer neural operator for handling non-uniform grids, which projects spatial points onto a small number of slices and facilitates information exchange between the slices. Chen et al. \cite{17} proposed a transformer neural operator that relies solely on coordinate representations. By eliminating the need for complex function values as inputs, this approach significantly improves efficiency.

In evaluating the effectiveness of a NO for turbulence simulation, in addition to comparing prediction accuracy over short time periods, another important criterion is its ability to achieve long-term stable predictions. Specifically, this involves assessing whether statistical quantities and structures remain consistent with high-precision numerical simulation results over extended periods. Li et al. \cite{18} showed that the FNO \cite{8} exhibited explosive behavior over time when predicting chaotic dissipative systems, and successfully predicted the long-term statistical behavior of dissipative chaotic systems by introducing dissipation regularization. Li et al. \cite{19} designed an implicit U-Net enhanced FNO (IU-FNO), which achieved accurate and stable long-term predictions in isotropic turbulence, free shear turbulence, and decaying turbulence. Oommen et al. \cite{20} applied a diffusion model to correct the long-term prediction results of the FNO, significantly improving the consistency between the predicted energy spectrum and the true distribution. 

However, most existing transformer neural operators tend to focus solely on reducing the single-step error and delaying the accumulation of errors in time integration as much as possible, while overlooking the ability to make long-term stable predictions. Li et al. \cite{21a} introduced the transformer-based neural operator (TNO), which combines a linear attention with spectral regression, achieving long-term stable predictions in both isotropic turbulence and free shear flow. However, the attention for each points often incurs significant computational overhead. Yang et al. \cite{21} fouced on the long-term prediction capability of low-rank transformer neural operators in turbulence simulation. In tests on three-dimensional isotropic turbulence, they showed that the original factorized transformer (FactFormer) \cite{15} did not diverge over extended periods, but the predicted high-wavenumber energy spectrum was inconsistent with the true value. By designing an implicit factorized transformer (IFactFormer), they successfully achieved long-term stable and accurate predictions. 

In this paper, we investigate the long-term prediction capability of the IFactFormer model for turbulent channel flows at three friction Reynolds numbers $\text{Re}_{\tau}\approx 180$, $395$, and $590$. We show that the original IFactFormer (IFactFormer-o) model in previous work \cite{21}, fails to achieve long-term stable predictions of turbulent channel flows. We identify potential causes for this failure and propose solutions. By making appropriate adjustments to the network architecture, specifically replacing the chained factorized attention with parallel factorized attention, we introduce a modified version, IFactFormer-m. While only adding minimal computational time and memory usage, the IFactFormer-m model significantly outperforms IFactFormer-o in terms of single-step prediction accuracy, and achieves precise long-term predictions of statistical quantities and flow structures. The IFactFormer-m model demonstrates more accurate and stable long-term predictions compared to neural operators including FNO [8], implicit FNO (IFNO) \cite{22}, and IFactFormer-o \cite{21}, as well as traditional LES models including the dynamic Smagorinsky model (DSM) and the wall-adapted local eddy-viscosity model (WALE). 

This paper consists of five sections. Section \ref{sec:2} presents the problem statement, introducing the Navier-Stokes (N-S) equations, the LES method, and the learning objectives of NO for this task. Section \ref{sec:3} presents several transformer neural operators and discusses modifications to the original IFactFormer model. Section \ref{sec:4} compares several machine learning models with traditional LES models in turbulent channel flows. Section \ref{sec:5} discusses from the perspective of attention kernels why IFactFormer-m converges faster and is more stable for long-term prediction compared to IFactFormer-o. Section \ref{sec:6} concludes the paper with a summary and points out the limitations of IFactFormer-m and possible improvements.

\section{Problem statement}\label{sec:2}
In this section, we first provide a brief introduction to the N-S equations and the LES method, followed by a discussion on the role of NO models and their learning objectives.

\subsection{Navier-Stokes equations}
Turbulence is widespread in nature and is a highly nonlinear, multiscale system. It is generally believed that its dynamics are governed by the N-S equations. The incompressible form of the N-S equations is as follows \cite{1}:
\begin{align}
&\frac{\partial u_i}{\partial x_i}=0, \label{eq:ns1} \\
&\frac{\partial u_i}{\partial t}+\frac{\partial\left(u_i u_j\right)}{\partial x_j}=-\frac{\partial p}{\partial x_i}+v \frac{\partial^2 u_i}{\partial x_j \partial x_j}+\mathcal{F}_i, \label{eq:ns2}
\end{align}
where $u_i$ denotes the velocity component along the $i$-th coordinate axis, $p$ represents the pressure normalized by the constant density $\rho$, $\nu$ is the kinematic viscosity, and $\mathcal{F}_i$ refers to the forcing term acting in the $i$-th direction. Consider the turbulent channel flows with the lower and upper walls located at $y=0$ and $y=2\delta$, respectively. 

Considering that the velocities in the three coordinate directions are $\left( u, v, w \right) = \left( u_1, u_2, u_3 \right)$ with fluctuations $\left( u', v', w' \right) = \left( u_1^\prime, u_2^\prime, u_3^\prime \right)$, the total shear stress is given by \cite{1}:
\begin{align}
\tau(y)=\rho \nu \frac{\partial\langle u\rangle}{\partial y}-\rho\left\langle u^{\prime} v^{\prime}\right\rangle,
\end{align}
where $\left\langle \cdot \right\rangle$ represents the spatial average over the homogeneous streamwise and spanwise directions, and $\left\langle u^{\prime} v^{\prime}\right\rangle$ is the Reynolds shear stress. At the wall, the boundary condition $u\left(\boldsymbol{x},t\right)=0$, so the wall shear stress is
\begin{align}
\tau_{\mathrm{w}} \equiv \rho \nu\left(\frac{\partial\langle u\rangle}{\partial y}\right)_{y=0}.
\end{align}
The friction velocity $u_\tau$ and viscous lengthscale $\delta_\nu$ are defined by \cite{1}:
\begin{align}
u_\tau \equiv \sqrt{\frac{\tau_{\mathrm{w}}}{\rho}}, \qquad \delta_v \equiv v \sqrt{\frac{\rho}{\tau_{\mathrm{w}}}}=\frac{v}{u_\tau}.
\end{align}
Therefore, the definition of the friction Reynolds number $\text{Re}_{\tau}$ is given by:
\begin{align}
\text{Re}_\tau \equiv \frac{u_\tau \delta}{\nu}=\frac{\delta}{\delta_\nu}.
\end{align}

\subsection{Large eddy simulation}
DNS involves directly solving the N-S equations on a fine mesh to fully resolve all the scales of the flow. However, the inherent multiscale nature of turbulence limits the applicability of DNS to turbulence problems at high Reynolds numbers due to its high computational cost. LES aims to resolve the large-scale turbulent structures on a coarse grid, thus reducing computational cost. Consider the spatial filtering operation as described below \cite{1}:
\begin{align}
\bar{f}(\boldsymbol{x})=\int_{\Omega} G(\boldsymbol{r}, \boldsymbol{x} ; \bar{\Delta}) f(\boldsymbol{x}-\boldsymbol{r}) d \boldsymbol{r}, \label{eq:filter}
\end{align}
where $G$ is the grid filter, $\bar{\Delta}$ is the filter width and $f$ is a physical quantity distributed over the spatial domain $\Omega$. The filtered N-S equations can be obtained by applying \Cref{eq:filter} to \Cref{eq:ns1} and \Cref{eq:ns2}, as follows \cite{1}: 
\begin{align}
&\frac{\partial \bar{u}_i}{\partial x_i}=0, \\
&\frac{\partial \bar{u}_i}{\partial t}+\frac{\partial\left(\bar{u}_i \bar{u}_j\right)}{\partial x_j}=-\frac{\partial \bar{p}}{\partial x_i}-\frac{\partial \tau_{i j}}{\partial x_j}+\nu  \frac{\partial^2 \bar{u}_i}{\partial x_j \partial x_j}+\overline{\mathcal{F}}_i.
\end{align}
Unlike the N-S equations, the filtered N-S equations are unclosed due to the introduction of unclosed SGS stress $\tau_{ij}$, which is defined as:
\begin{align}
\tau_{i j}=\overline{u_i u_j}-\bar{u}_i \bar{u}_j .
\end{align}

The dynamic Smagorinsky model (DSM) \cite{24} and wall-adapting local eddy-viscosity (WALE) model \cite{27} are traditional LES methods that both account for wall effects in their modeling of SGS stresses.

\subsection{Problem definition}
In this study, our objective is to develop a neural operator trained on coarse grids with large time steps, aiming to learn the dynamics of large-scale turbulent structures which are obtained by a low-pass filter. Considering that $\mathcal{U}_t$ is a Banach space of filtered turbulent velocity field $\bar{u}$ depending on time $t$, defined on compact domains $\mathcal{X} \subset \mathbb{R}^{3}$, mapping into $\mathbb{R}^{3}$. The true operator is denoted by $\mathcal{H}: \mathcal{U}_t \rightarrow \mathcal{U}_{t+\delta t}$, where $\delta t$ is the time step. The goal of neural operator is to develop a model $\hat{\mathcal{H}}_{\phi}$ parameterized by $\phi \in \Phi$, where the optimal parameters $\phi^*$ are identified by solving the minimization problem as follows \cite{7}:
\begin{align}
\min_{\phi \in \Phi} \sum_{j=1}^{M} \sum_{i=1}^{N} \left\| \hat{\mathcal{H}}_{\phi}\left[\bar{\mathbf{u}}^{(t)}_j\right]\left(x_i\right) - \bar{\mathbf{u}}^{(t+\delta t)}_j\left(x_i\right) \right\|_2,
\end{align}
thereby approximating the true operator $\mathcal{H}$. Here, $M$ represents the number of input–output pairs. The vectors $\bar{\mathbf{u}}^{(t)} = \left[ \bar{u}\left(t,x_1\right), \bar{u}\left(t,x_2\right), \ldots, \bar{u}\left(t,x_N\right) \right]$ and $\bar{\mathbf{u}}^{(t+\delta t)} = \left[ \bar{u}\left(t+\delta t,x_1\right), \bar{u}\left(t+\delta t,x_2\right), \ldots, \bar{u}\left(t+\delta t,x_N\right) \right]$ correspond to a function $\bar{u}^{(t)} \in \mathcal{U}_t$ and $\bar{u}^{(t+ \delta t)} \in \mathcal{U}_{t+\delta t}$ evaluated at a set of fixed locations $\left\{ x_i \right\}_{i=1}^{N} \subset \mathcal{X}$, respectively.

\section{Transformer neural operator}\label{sec:3}
In this section, the first part explains how attention-based neural operators approximate the true operator by leveraging a parameterized integral transform. The second part discusses the drawback of the chained factorized attention, and proposes the parallel factorized attention. The final part introduces the overall architecture of the modified implicit factorized transformer. 

\subsection{Attention-based integral neural operator}
\begin{figure*}[t]
\centering
\includegraphics[scale=0.6]{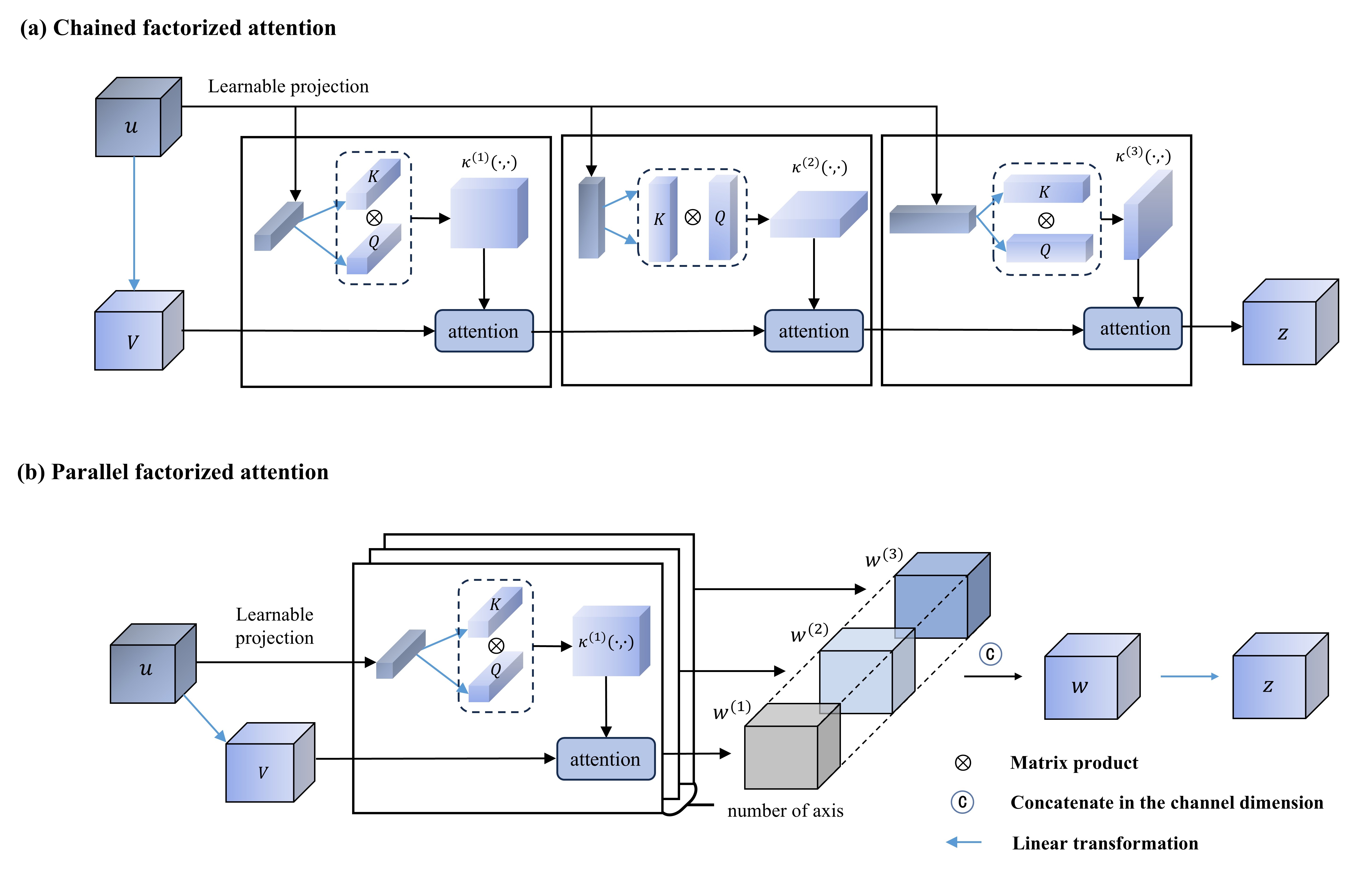}
\caption{(a) The original factorized attention, which processes each axis sequentially; (b) The modified factorized attention, which processes each axis in parallel.}
\label{figure:fact-attn}
\end{figure*}

The self-attention mechanism \cite{28} dynamically weights the input by computing the correlations between different positions in the input vector, thereby capturing the dependencies among various positions. The perspectives of previous work \cite{16,29,30} demonstrated that the standard attention mechanism can be viewed as a Monte Carlo approximation of an integral operator. Considering an input vector $\mathbf{u}_i \in \mathbb{R}^{1 \times d_{\text {in}}}$ with $d_{\text {in}}$ channels in $N$ points ($1 \le i \le N$), query $\mathbf{q}_i \in \mathbb{R}^{1 \times d}$, key $\mathbf{k}_i \in \mathbb{R}^{1 \times d}$ and value $\mathbf{v}_i \in \mathbb{R}^{1 \times d}$ vectors with $d$ channels are first generated through linear transformations as follows:
\begin{align} \label{eq:attn1}
\mathbf{q}_i = \mathbf{u}_i \mathbf{W_q}, \
\mathbf{k}_i = \mathbf{u}_i \mathbf{W_k}, \
\mathbf{v}_i = \mathbf{u}_i \mathbf{W_v},  
\end{align}
where $\left \{ \mathbf{W_q}, \mathbf{W_k}, \mathbf{W_v} \right \} \in \mathbb{R}^{d_{in} \times d}$. Subsequently, the attention weights $\alpha_{i j}$ are computed for $\mathbf{q}$ and $\mathbf{k}$ using the following equation:
\begin{align}  \label{eq:attn2}
\alpha_{i j}=\frac{\exp \left[g\left(\mathbf{q}_i, \mathbf{k}_j\right)\right]}{\sum_{s=1}^n \exp \left[g\left(\mathbf{q}_i, \mathbf{k}_s\right)\right]},
\end{align}
where $g$ is a scaled dot-product as follows:
\begin{align}
g\left(\mathbf{q}_i, \mathbf{k}_j\right) = \frac{\mathbf{q}_i \cdot \mathbf{k}_j}{\sqrt{d}}.
\end{align}
Finally, the attention weights $\alpha_{i j}$ are applied to the value vectors $\mathbf{v}_j$ to capture the dependencies between different positions in the sequence as follows:
\begin{align}
\mathbf{z}_i=\sum_{j=1}^n \alpha_{i j} \mathbf{v}_j \approx \int_{\Omega} \kappa \left ( x_i, \psi \right )v_j \left ( \psi  \right )d \psi. \label{eq:kernal}
\end{align}
Here, the $i$-th row vector $\left(\alpha^i\right)_j$ is regarded as the global kernel function $\kappa \left ( x_i, \psi \right )$ of the approximate integral operator for point $x_i$. 

\subsection{Factorized attention}
The standard self-attention mechanism is often criticized for its quadratic computational complexity. Li et al. \cite{15} proposed the factorized attention, which alleviates this issue. Due to the need for chained integration along each axis, we refer to this as the chained factorized attention, as illustrated in \cref{figure:fact-attn}(a). Specifically, for points on a Cartesian grid with $N_1 \times N_2 \times N_3 = N$ points in an three-dimensional space $\Omega_1 \times \Omega_2 \times \Omega_3$, the chained factorized attention decomposes the kernel function in \Cref{eq:kernal} into three separate kernel functions along each axis $\left\{\kappa^{(1)}, \kappa^{(2)}, \kappa^{(3)}\right\}: \mathbb{R} \times \mathbb{R} \mapsto \mathbb{R}$, and performs the integral transformation in the following manner \cite{15}:
\begin{small}
\begin{align}
z\left(x_{i_1}^{(1)}, x_{i_2}^{(2)}, x_{i_3}^{(3)}\right)= & \int_{\Omega_3} \kappa^{(3)}\left(x_{i_3}^{(3)}, \psi_3\right) \int_{\Omega_{2}} \kappa^{(2)}\left(x_{i_{2}}^{(2)}, \psi_{2}\right) \nonumber \\
 &\int_{\Omega_1} \kappa^{(1)}\left(x_{i_1}^{(1)}, \psi_1\right) v\left(\psi_1, \psi_2, \psi_3\right) d\psi_1 d\psi_2 d\psi_3, \label{eq:int_trans}
\end{align} 
\end{small}where each kernel is obtained through a learnable projection followed by \Cref{eq:attn1} and \Cref{eq:attn2}. The goal of the learnable projection is to compress the original input vector onto each axis. The calculation formulas for the three axes are as follows:
\begin{align}
\phi^{(1)}\left ( x_{i_1}^{(1)} \right ) = h^{(1)} \omega^{(1)} \int_{\Omega_2} \int_{\Omega_3} \gamma^{(1)} u\left ( x_{i_1}^{(1)}, \psi_2, \psi_3 \right )d\psi_2 d\psi_3, \\  
\phi^{(2)}\left ( x_{i_2}^{(2)} \right ) = h^{(2)} \omega^{(2)} \int_{\Omega_1} \int_{\Omega_3} \gamma^{(2)} u\left ( \psi_1, x_{i_2}^{(2)}, \psi_3 \right )d\psi_1 d\psi_3, \\
\phi^{(3)}\left ( x_{i_3}^{(3)} \right ) = h^{(3)} \omega^{(3)} \int_{\Omega_1} \int_{\Omega_2} \gamma^{(3)} u\left ( \psi_1, \psi_2, x_{i_3}^{(3)} \right )d\psi_1 d\psi_2.
\end{align}
Here, $\omega^{(s)} = N_s / N$ is a constant, $\left \{  h^{(1)},  h^{(2)},  h^{(3)} \right \}$ are multilayer perceptron (MLP) and $\left \{  \gamma^{(1)},  \gamma^{(2)},  \gamma^{(3)} \right \}$ are linear transformation.

The above scheme has a drawback: all kernel functions $\left\{\kappa^{(1)}, \kappa^{(2)}, \kappa^{(3)}\right\}$ are derived from the original input function $\boldsymbol{u}$. These kernel functions are often effective at capturing the dependencies between different positions within the current function. However, for $\kappa^{(2)}$ and $\kappa^{(3)}$, the dependencies need to be evaluated on two new function 
\begin{small}
\begin{align}
&\int_{\Omega_1} \kappa^{(1)}\left(x_{i_1}^{(1)}, \psi_1\right) v\left(\psi_1, \psi_2, \psi_3\right) d\psi_1,  \\
&\int_{\Omega_2} \kappa^{(2)}\left(x_{i_2}^{(2)}, \psi_2\right) \int_{\Omega_1} \kappa^{(1)}\left(x_{i_1}^{(1)}, \psi_1\right) v\left(\psi_1, \psi_2, \psi_3\right) d\psi_1 d\psi_2. 
\end{align}
\end{small}Considering that these two functions mentioned above are obtained through a series of complex computations involving the input function $\boldsymbol{u}$ and parameters of neural network, and since the parameters are unknown to the kernel functions $\kappa^{(2)}$ and $\kappa^{(3)}$, they are tasked with evaluating the dependencies on an unknown system. This undoubtedly presents a significant challenge. In the works of Li et al. \cite{15} and Yang et al. \cite{21}, the FactFormer and IFactFormer models, which are based on chained factorized attention, achieved promising test results in certain two-dimensional flows and three-dimensional isotropic turbulence. This success is likely due to the fact that, for isotropic problems, the learned dependencies in different axes are consistent, making the evaluation relatively easier. This, to some extent, obscures the underlying issue.

Based on the above analysis, we propose parallel factorized attention, as illustrated in \cref{figure:fact-attn}(b). Compared to \Cref{eq:int_trans}, the form of the integral transformation is modified as follows:
\begin{align}
&w^{(s)}= \int_{\Omega_s} \kappa^{(s)}\left(x_{i_s}^{(s)}, \psi_s\right) v\left(\psi_1, \psi_2, \psi_3\right) d\psi_s, \\
&w = \text{Concat}\left ( w^{(1)}, w^{(2)}, w^{(3)} \right ), \label{eq:concat} \\
&z= \text{Linear}\left ( w\right ) \label{eq:linear}.
\end{align}
Here, $s=1, 2, 3 $, ``$\text{Linear}$'' is a linear transformation from $\mathbb{R}^{3d}$ to $\mathbb{R}^{d}$, and ``$\text{Concat}$'' means that concatenate the input function on channel dimension. The above simple modification allows each kernel function to focus on learning the dependencies along different axes of the current function.

\subsection{Implicit factorized transformer}
\begin{figure}[H]
\centering
\includegraphics[scale=0.6]{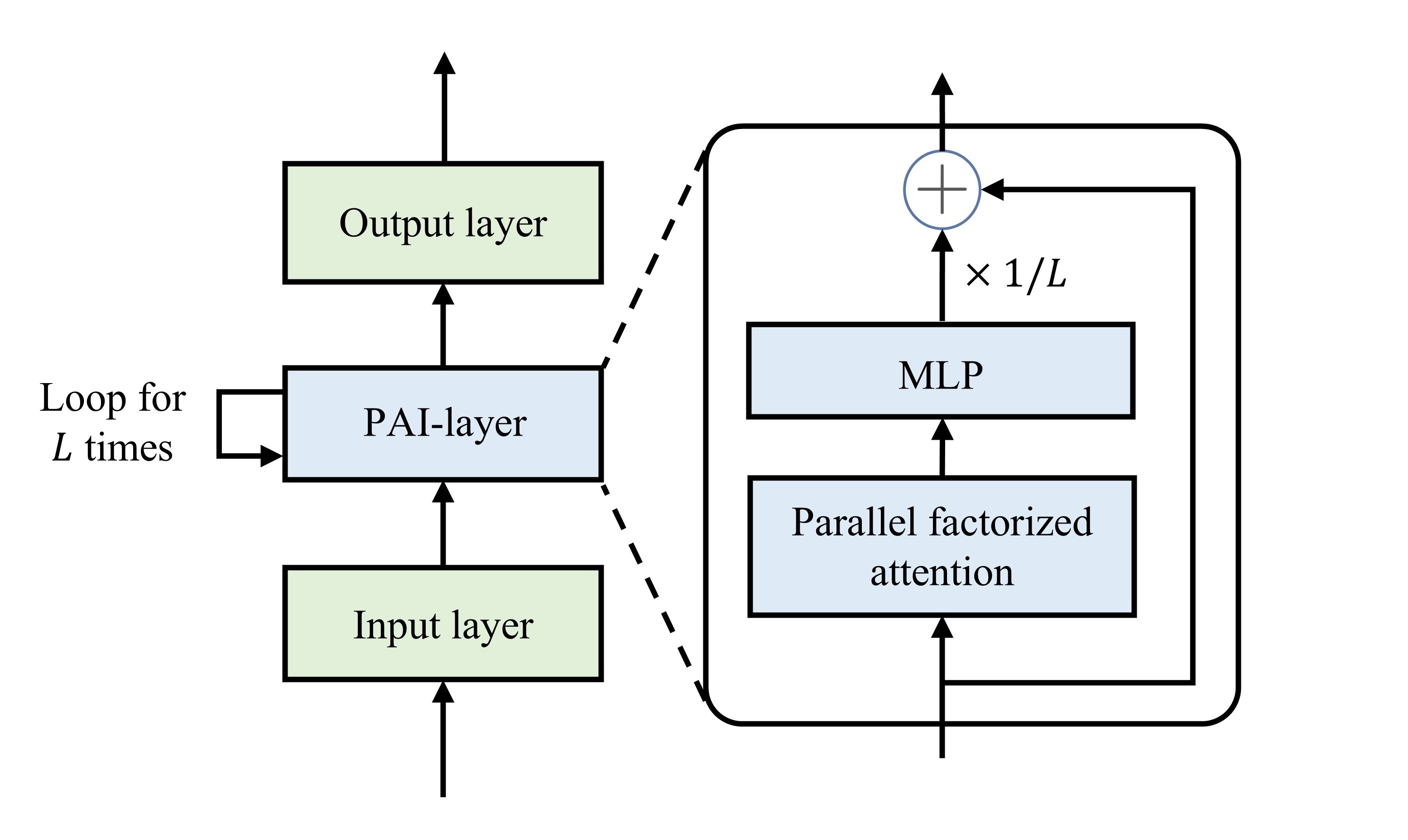}
\caption{Overall design of IFactFormer-m. The left side presents the overall framework based on implicit iteration. The right side illustrates the internal structure of the parallel axial integration layer (PAI-layer).} \label{figure:IFactFormer-m}
\end{figure}

By utilizing the designed parallel factorized attention, we propose the modified implicit factorized transformer model (IFactFormer-m), as illustrated in \cref{figure:IFactFormer-m}. The IFactFormer-m model consists of three components: the input layer $\mathcal{I}: \mathbb{R}^{d_{in}} \to \mathbb{R}^{d}$, parallel axial integration layer (PAI-layer) $\mathcal{P}: \mathbb{R}^{d} \to \mathbb{R}^{d}$, and the output layer $\mathcal{O}: \mathbb{R}^{d} \to \mathbb{R}^{d_{out}}$. For the temporal prediction of three-dimensional incompressible turbulence, it is assumed that $d_{in} = d_{out} = 3$. The input and output layers are three-layer MLPs, used to map the input function to a high-dimensional space and project the function from the high-dimensional space back to the low-dimensional space. The PAI-layer is a global nonlinear operator, used to approximate the integral transformation in the high-dimensional space to update the input function. Consistent with previous works \cite{19,21,22,31}, we adopt an implicit iteration strategy, where the parameters are shared across each PAI-layer. This approach effectively enhances the stability of the model in long-term turbulent flow predictions. Therefore, the overall operator of the $L$-layer IFactFormer-m model can be expressed as $\mathcal{O} \circ \underbrace{\mathcal{P} \circ \cdots  \circ \mathcal{P}}_{L} \circ \mathcal{I}$. 

Consider the input function $u \in \mathbb{R}^{d}$ discretized into the vector $\mathbf{u}_i \in \mathbb{R}^{1 \times d}$ at points $\left \{x_i \right \}_{i=1}^{N}$. The computation formula for the PAI-layer is given as follows:
\begin{align}
\mathbf{u}_i^{'} &= \mathbf{u}_i + \frac{1}{L} \text{MLP}\left ( \mathbf{z}_i \right ) \nonumber\\
&= \mathbf{u}_i + \frac{1}{L} \text{MLP}\left ( \text{P-Fact-Attn}\left ( \mathbf{u}_i \right ) \right ),
\end{align}
where ``$\text{MLP}$'' is three-layer, ``$\text{P-Fact-Attn}$'' denotes the parallel factorized attention and the output vector $\mathbf{u}_i^{'}$ is the discretized representation of the updated function $u'$. The factor of $1/L$ performs scale compression, ensuring that the final scale remains consistent for any given number of implicit iterations $L$.

\section{Numerical Results}\label{sec:4}
In this section, the first part discusses the construction of the dataset of turbulent channel flows. The second part compares the IFactFormer-m with other ML models, including FNO, IFNO and IFactFormer-o, as well as traditional LES methods including DSM and WALE.

\begin{table*}[!t]
\footnotesize
\caption{Parameters for the DNS, LES and ML of turbulent channel flow.}
\tabcolsep 16pt
\begin{tabular*}{\textwidth}{cccccccc}
\toprule
& Resolution & Domain & $\text{Re}_{\tau}$ & $\nu$ & $\Delta X^{+}$ & $\Delta Y^{+}_w$ & $\Delta Z^{+}$ \\ 
\hline
\multirow{3}{*}{DNS} & $192 \times 129 \times 64$ & $\left [ 4\pi, 2, 4\pi /3 \right ]$ & $180$ & $1/4200$ & $11.6$ & $0.98$ & $11.6$ \\
& $256 \times 193 \times 128$ & $\left [ 4\pi, 2, 4\pi /3 \right ]$ & $395$ & $1/10500$ & $19.1$ & $1.4$ & $12.8$ \\
& $384 \times 257 \times 192$ & $\left [ 4\pi, 2, 4\pi /3 \right ]$ & $590$ & $1/16800$ & $19.3$ & $1.6$ & $12.9$ \\
\hline
\multirow{3}{*}{LES \& ML} & $32 \times 33 \times 16$ & $\left [ 4\pi, 2, 4\pi /3 \right ]$ & $180$ & $1/4200$ & $69.6$ & $3.93$ & $46.4$ \\
& $64 \times 49 \times 32$ & $\left [ 4\pi, 2, 4\pi /3 \right ]$ & $395$ & $1/10500$ & $76.4$ & $5.6$ & $51.2$ \\
& $64 \times 65 \times 32$ & $\left [ 4\pi, 2, 4\pi /3 \right ]$ & $590$ & $1/16800$ & $115.8$ & $6.4$ & $77.4$ \\
\bottomrule
\label{tab:param}
\end{tabular*}
\end{table*}

\begin{table*}[!t]
\footnotesize
\caption{Model configurations of FNO, IFNO, IFactFormer-o and IFactFormer-m.}
\tabcolsep 30pt
\begin{tabular*}{\textwidth}{lccc}
\toprule
Model & $\text{Re}_{\tau} \approx 180$ & $\text{Re}_{\tau} \approx 395$ & $\text{Re}_{\tau} \approx 590$ \\ 
\hline
FNO \& IFNO & \makecell[l]{Layer: 10 \\ Modes: 8 \\ Dim: 96} & \makecell[l]{Layer: 5 \\ Modes: 16 \\ Dim: 64} & \makecell[l]{Layer: 5 \\ Modes: 16 \\ Dim: 64} \\
\hline
IFactFormer-o \& IFactFormer-m & \makecell[l]{Layer: 10 \\ Heads: 5 \\ Dim: 96} & \makecell[l]{Layer: 5 \\ Heads: 5 \\ Dim: 96} & \makecell[l]{Layer: 5 \\ Heads: 5 \\ Dim: 96} \\
\bottomrule
\label{tab:param_model}
\end{tabular*}
\end{table*}

\begin{table*}[!t]
\footnotesize
\caption{The training costs of each individual epoch for the four FNO, IFNO, IFactFormer-o and IFactFormer-m.}
\tabcolsep 15.1pt
\begin{tabular*}{\textwidth}{lcccccc}
\toprule
\multirow{2}{*}{Model} & \multicolumn{2}{c}{$\text{Re}_{\tau} \approx 180$} & \multicolumn{2}{c}{$\text{Re}_{\tau} \approx 395$} & \multicolumn{2}{c}{$\text{Re}_{\tau} \approx 590$} \\ 
\cline{2-7}
 & Time(s) & Memory(GB) & Time(s) & Memory(GB) & Time(s) & Memory(GB) \\
\hline
FNO & 1154 & 14.8 & 2432 & 29.9 & 2905 & 31.5 \\
IFNO & 785 & 4.3 & 1851 & 13.4 & 2325 & 14.7 \\
IFactFormer-o & 513 & 11.5 & 2585 & 13.8 & 3301 & 18.1 \\
IFactFormer-m & 690 & 19.2 & 3118 & 23.8 & 4007 & 31.3 \\
\bottomrule
\label{tab:train_time}
\end{tabular*}
\end{table*}

\subsection{Dataset of turbulent channel flows}
We compute turbulent channel flows at three friction Reynolds numbers $\text{Re}_{\tau}\approx$ $180$, $395$, and $590$ on fine grids using DNS, employing an open-source framework Xcompact3D \cite{32,33}. We perform LES calculations and train ML models on a coarse grid. \cref{tab:param} presents the relevant parameters, with all simulations conducted in a cuboid domain of size $\left [ 4\pi, 2, 4\pi /3 \right ]$, where the X-direction represents the streamwise direction, the Y-direction the wall-normal direction, and the Z-direction the spanwise direction. $\Delta X^{+}$ and $\Delta Z^{+}$ represent the normalized grid spacings in the streamwise and spanwise directions, respectively, while $\Delta Y^{+}_w$ indicates the distance from the wall to the first grid point. The superscript ``+'' denotes a distance that has been non-dimensionalized by the viscous lengthscale $\delta_{\nu}$, e.g., $y^{+} = y / \delta_{\nu}$. 

We perform filtering and interpolation on the DNS data to obtain filtered DNS (fDNS) data on the LES grid. The fDNS data is then used for training and testing the ML models. The DNS time step is set to 0.005, while the time step of the ML model is 200 times larger. Considering the wall viscous time $\tau_{\nu} = \delta_{\nu}^2 / \nu$, the wall viscous time steps of the ML model are $7.5 \tau_{\nu}$, $14.6 \tau_{\nu}$, and $20.7 \tau_{\nu}$ for the three Reynolds numbers $\text{Re}_{\tau} \approx$ $180$, $395$, and $590$, respectively. A total of 21 fDNS datasets are generated, each retaining data from 400 snapshots. Among the first 20 datasets, 80\% and 20\% are randomly selected for training and testing the model, respectively, with the final dataset reserved for post-analysis. For all ML models, the previous frame snapshot of velocity field $u^{(T)}$ is used to predict the next frame snapshot $u^{(T+1)}$.

The first comparison focuses on several different ML models, with an emphasis on both the accuracy of short-term predictions and the stability of long-term predictions. To ensure a fair comparison, the training parameters for all models are kept consistent across the same dataset. The AdamW optimizer \cite{34} is employed, with an initial learning rate of 0.0005. The step learning rate scheduler is multiplied by a factor of 0.7 every 5 epochs, and the batch size is 2. Detailed hyperparameter settings for various models are provided in \cref{tab:param_model}. Here, ``Layer'' refers to the number of layers for the FNO model, while for the other three models, it represents the number of implicit iterations. ``Modes'' refers to the number of frequencies retained in the frequency domain, ``Heads'' refers to the number of multi-head attention and ``Dim'' refers to the number of channels in the latent space. 

\cref{tab:train_time} presents the training overhead of various models, including computation time and memory usage. IFactFormer-m incurs additional computational overhead compared to IFactFormer-o. The increase in memory usage stems from the need to store more intermediate variables simultaneously in \Cref{eq:concat}, while the rise in computational load is due to the input dimension of the linear layer in \Cref{eq:linear} being three times larger than before. Therefore, properly reducing the number of channels in the latent space while increasing the number of parameters elsewhere can reduce the additional computational of IFactFormer-m.

\subsection{Results comparison}
\begin{figure*}[!t]
\centering
\includegraphics[scale=1]{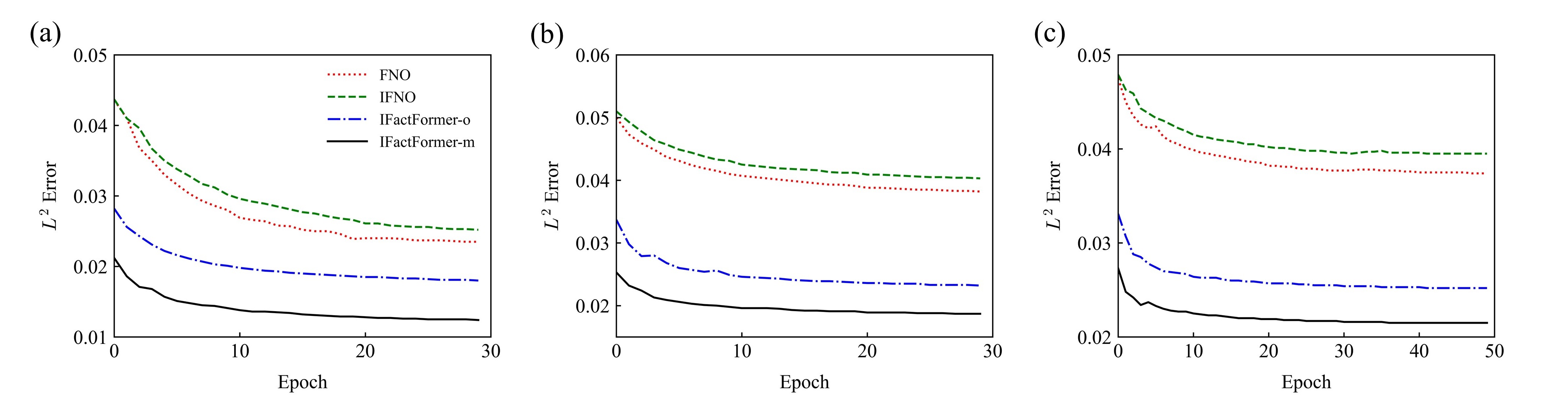}
\caption{The test loss of FNO, IFNO, IFactFormer-o and IFactFormer-m models at various Reynolds numbers: (a) $\text{Re}_{\tau}\approx 180$; (b) $\text{Re}_{\tau}\approx 395$; (c) $\text{Re}_{\tau}\approx 590$. Note that the test loss of different models when untrained is omitted here.}
\label{fig:loss}
\end{figure*}

\begin{figure*}[!t]
\centering
\includegraphics[scale=1]{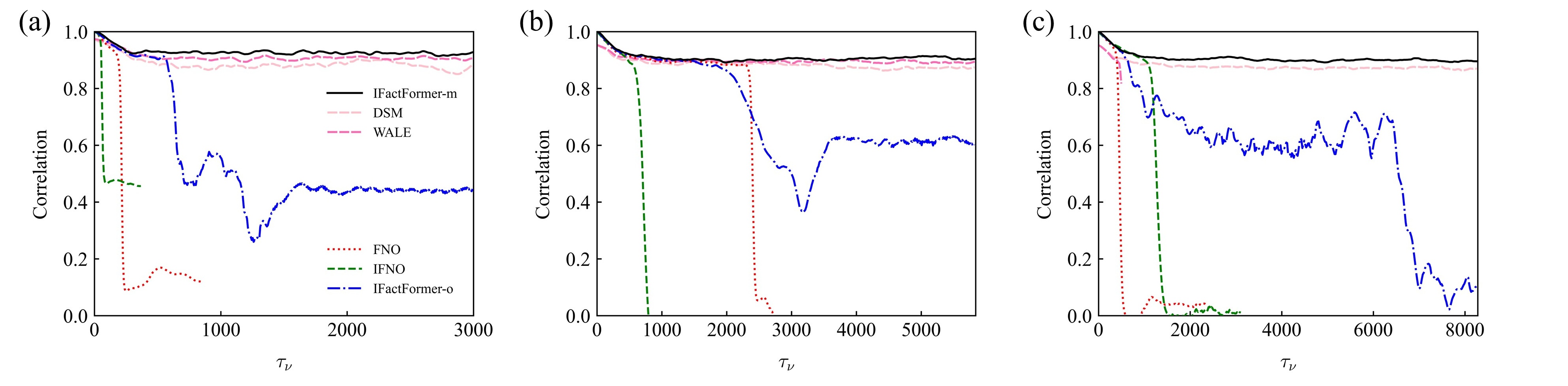}
\caption{The correlation coefficient curve of streamwise velocity using FNO, IFNO, IFactFormer-o and IFactFormer-m models at various Reynolds numbers: (a) $\text{Re}_{\tau}\approx 180$; (b) $\text{Re}_{\tau}\approx 395$; (c) $\text{Re}_{\tau}\approx 590$.}
\label{fig:cor}
\end{figure*}

\begin{figure*}[!t]
\centering
\includegraphics[scale=0.28]{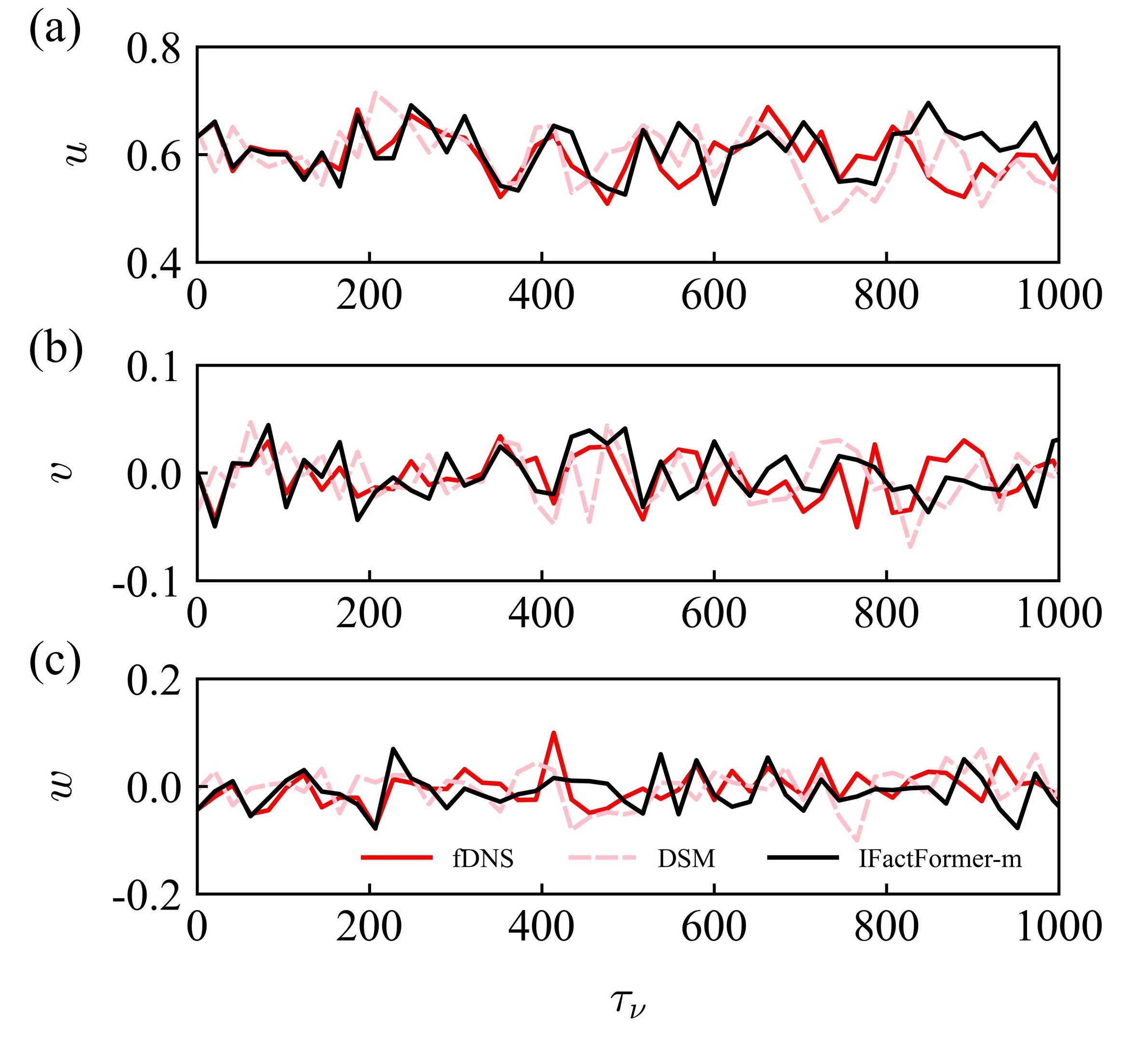}
\caption{The time series of velocity at position $[2\pi,0.27,2\pi/3]$ at $\text{Re}_{\tau}\approx 590$: (a) streamwise velocity; (b) wall-normal velocity; (c) spanwise velocity.}
\label{fig:point_time_Re590}
\end{figure*}

\begin{figure*}[!b]
\centering
\includegraphics[scale=0.85]{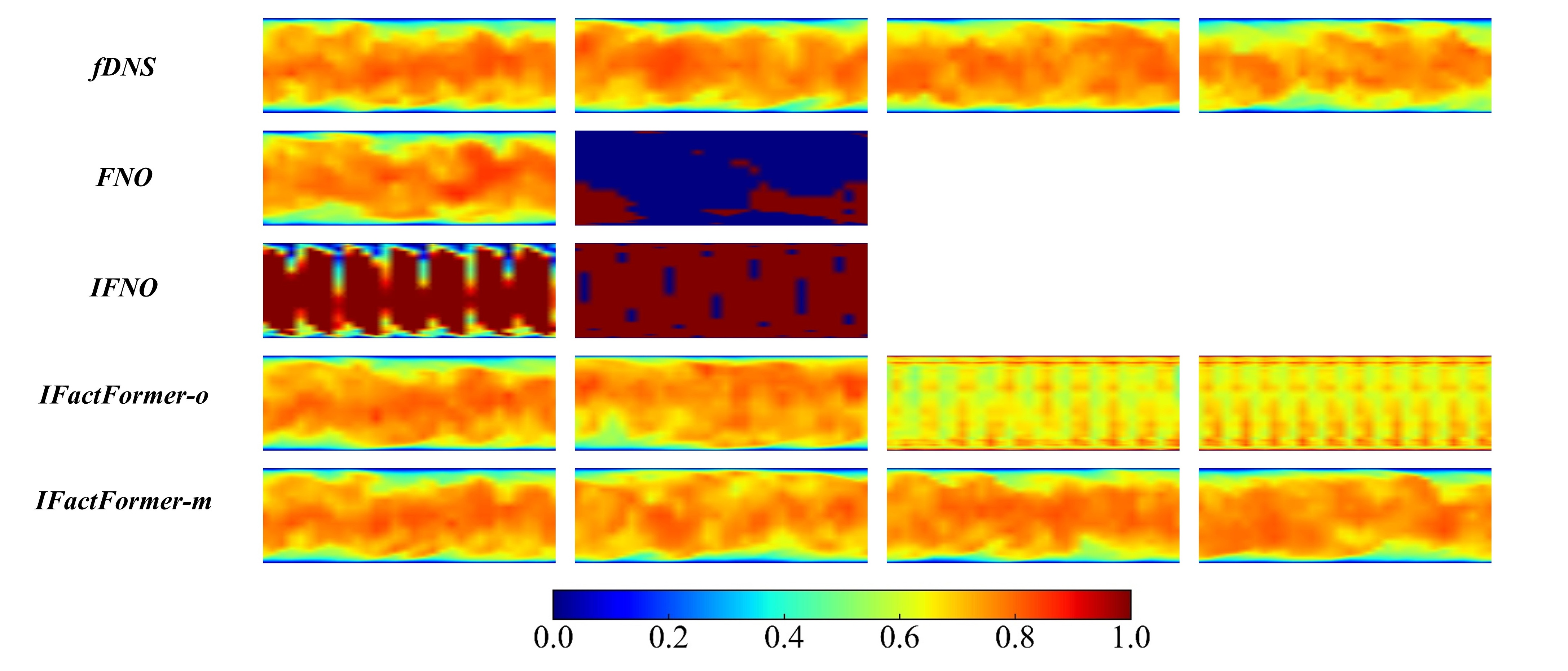}
\caption{Evolution of streamwise velocity (in an x-y plane) for the turbulent channel flow at $\text{Re}_{\tau}\approx 180$. From left to right, the snapshots correspond to the 10th, 50th, 200th, and 400th time steps, respectively.}
\label{figure:vel_180}
\end{figure*}

The relative $L_2$ error is used as the loss function for both training and testing as follows:
\begin{align}
 L_2 =\frac{\| \hat{u} - u \|_2}{\| u \|_2},
\end{align}
where $\hat{u}$ represents the predicted velocity field and $u$ is the ground truth of velocity field.

\Cref{fig:loss} shows the test loss curves for four ML models in turbulent channel flows with different friction Reynolds numbers $\text{Re}_{\tau}$. Here, the test loss of the models when untrained is omitted, at which point the test loss for all models is approximately 1. Both the IFactFormer-o and IFactFormer-m models outperform FNO and IFNO in terms of convergence speed and accuracy. The IFactFormer-m model achieves a higher accuracy after just one training step, surpassing the accuracy of FNO and IFNO at convergence. This result demonstrates the powerful fitting capability of the transformer-based model, enabling high-precision predictions in a short time frame. The accuracy of IFactFormer-m is significantly higher than that of IFactFormer-o, which demonstrates the effectiveness of the parallel factorized attention. Additionally, we observe that as the Reynolds number increases, the test error at convergence for all models tends to rise, indicating that the learning difficulty increases with the growing nonlinearity of the system.

We utilize a group of data that is excluded from both the training and test sets, and perform long-term forecasting using four different ML models through an autoregressive approach. The total number of forecasted time steps is 400, which means that the fluid passes through the channel approximately 21.25 times in a physical sense. The total wall viscous time spans are $3000 \tau_{\nu}$, $5840 \tau_{\nu}$, and $8280 \tau_{\nu}$ for the three Reynolds numbers, respectively. By analyzing these predictions, we can compare the long-term forecasting capabilities of the different models.

Pearson correlation coefficient is used to measure the degree of linear correlation between two variables $a$ and $b$, and its formula is given as follows:
\begin{align}
r=\frac{\sum_{i=1}^n\left(a_i-\bar{a}\right)\left(b_i-\bar{b}\right)}{\sqrt{\sum_{i=1}^n\left(a_i-\bar{a}\right)^2 \sum_{i=1}^n\left(b_i-\bar{b}\right)^2}}.
\end{align}
Here $n$ is the number of grids, $\bar{(\cdot)}$ representes the mean values over the spatial grid. This coefficient closer to 1 indicates a stronger correlation. \Cref{fig:cor} shows the Pearson correlation coefficients among the predicted streamwise velocity from four models, two LES methods, and fDNS at each forecasted time step. The correlation of FNO and IFNO sharply declines within very short time steps, even dropping below zero at $\text{Re}_{\tau}\approx$ $395$, and $590$, followed by divergent behavior that prevents further predictions in these three cases. Although the IFactFormer-o model is capable of making 400-step predictions, the correlation coefficient gradually decreases over time. Among the four ML models, only the IFactFormer-m model is able to maintain a correlation coefficient around 0.9 over the 400-step prediction. In the traditional LES methods, the WALE model diverges at $\text{Re}_{\tau}\approx 590$. For the other two Reynolds numbers $\text{Re}_{\tau}\approx$ $180$, and $395$, the correlation coefficients of the WALE model surpass those of the DSM model, but are slightly lower than those of the IFactFormer-m model. \Cref{fig:point_time_Re590} presents the time series of velocity from 0 to $1000 \tau_{\nu}$ at position $[2\pi,0.27,2\pi/3]$ at $\text{Re}_{\tau}\approx 590$. The IFactFormer-m curves exhibit a better agreement with fDNS profiles compared to DSM.

\begin{figure*}[!t]
\centering
\includegraphics[scale=0.85]{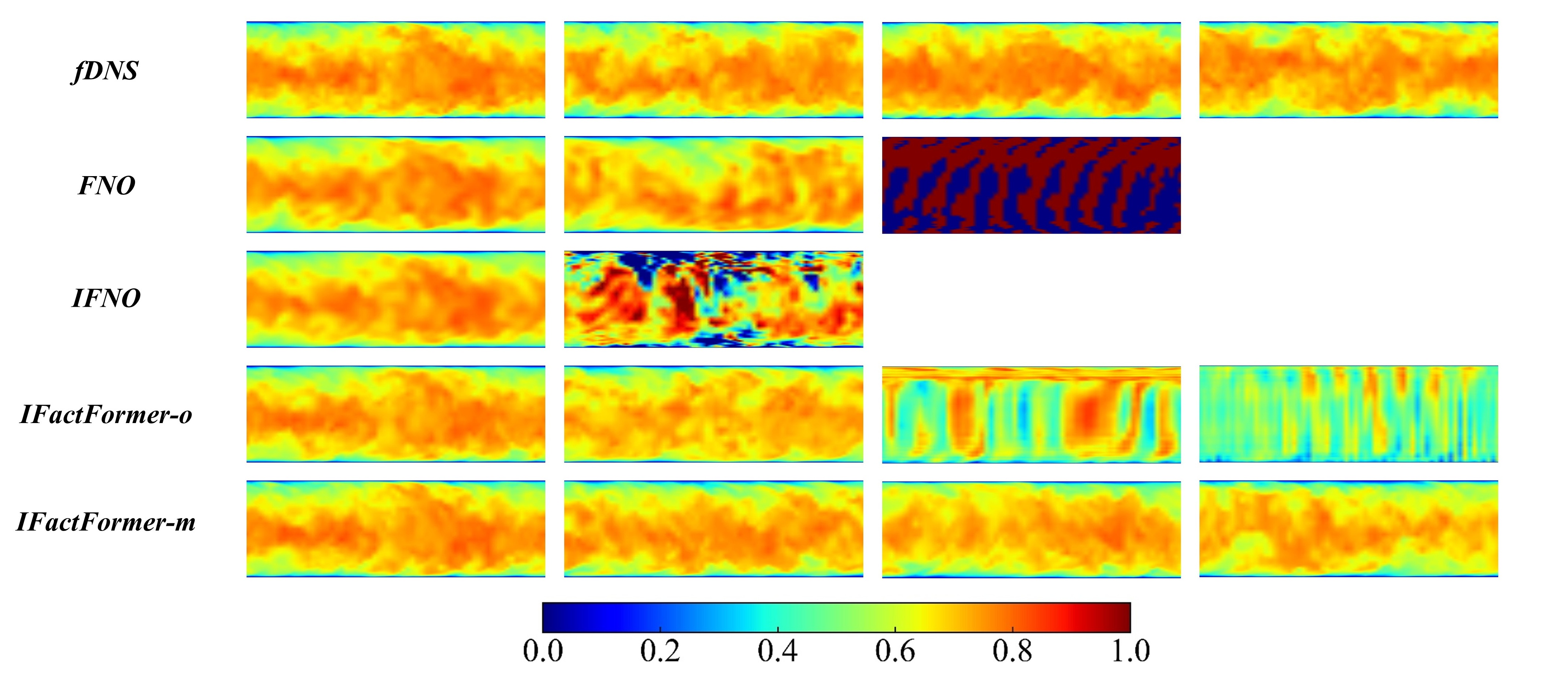}
\caption{Evolution of streamwise velocity (in an x-y plane) for the turbulent channel flow at $\text{Re}_{\tau}\approx 395$. From left to right, the snapshots correspond to the 10th, 50th, 200th, and 400th time steps, respectively.}
\label{figure:vel_395}
\end{figure*}

\begin{figure*}[!t]
\centering
\includegraphics[scale=0.85]{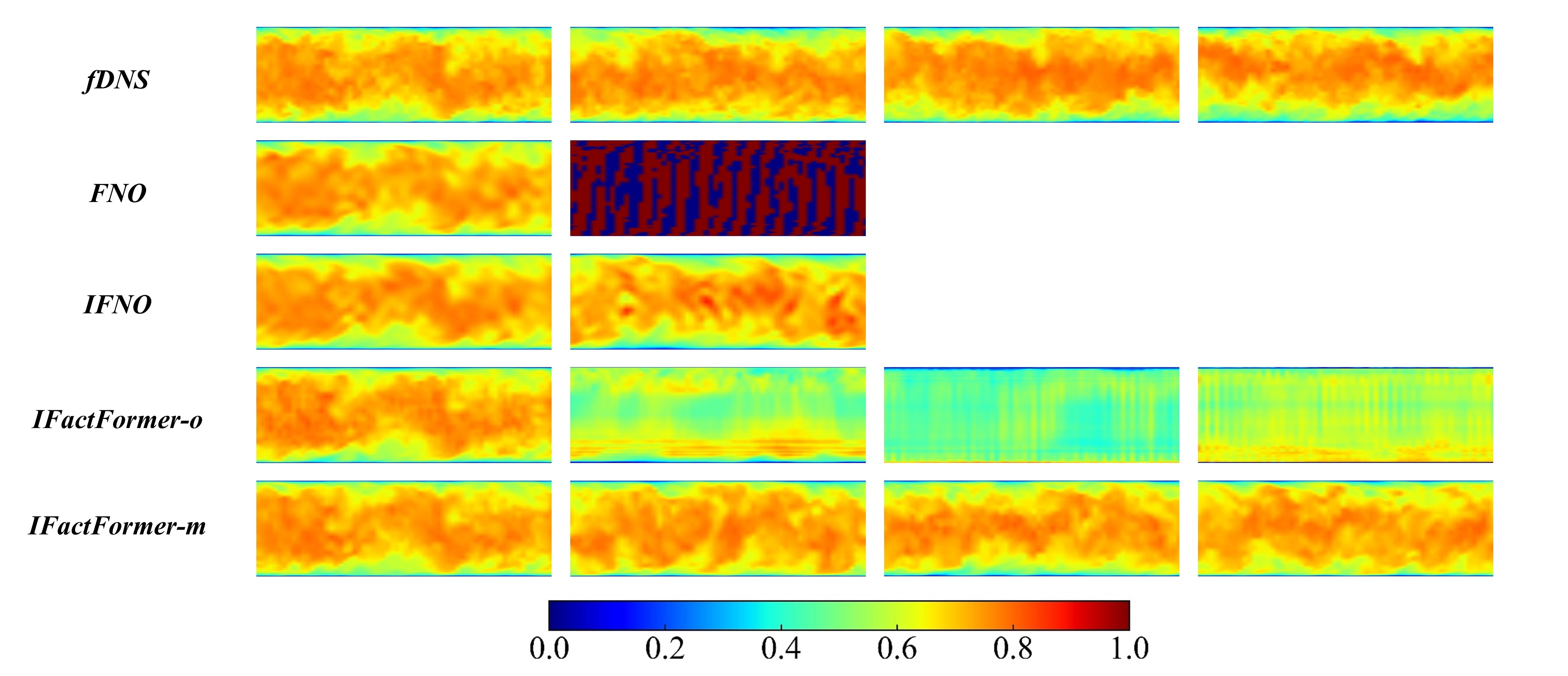}
\caption{Evolution of streamwise velocity (in an x-y plane) for the turbulent channel flow at $\text{Re}_{\tau}\approx 590$. From left to right, the snapshots correspond to the 10th, 50th, 200th, and 400th time steps, respectively.}
\label{figure:vel_590}
\end{figure*}

\Cref{figure:vel_180}-\ref{figure:vel_590} present cross-sectional snapshots of the streamwise velocity fields in an x-y plane predicted by the four ML models at different Reynolds numbers in the 10th, 50th, 200th, and 400th time steps. As the Reynolds number increases, the turbulent channel flow exhibits more small-scale features. Comparing the images in the first column of \Cref{figure:vel_395} and \ref{figure:vel_590}, it can be observed that FNO and IFNO have lost a significant amount of small-scale structures in their predictions at the tenth time step, only capturing the relatively larger-scale structures. In contrast, both IFactFormer-o and IFactFormer-m are able to retain the small-scale structures. This may be attributed to the high-frequency truncation in the frequency domain required by FNO and IFNO. As time progresses, the IFactFormer-o model begins to predict an increasing number of ``non-physical'' states. In contrast, the IFactFormer-m model significantly alleviates this issue, accurately maintaining multi-scale structures of turbulent channel flows even after 400 time steps.

\begin{figure*}[!t]
\centering
\includegraphics[scale=1]{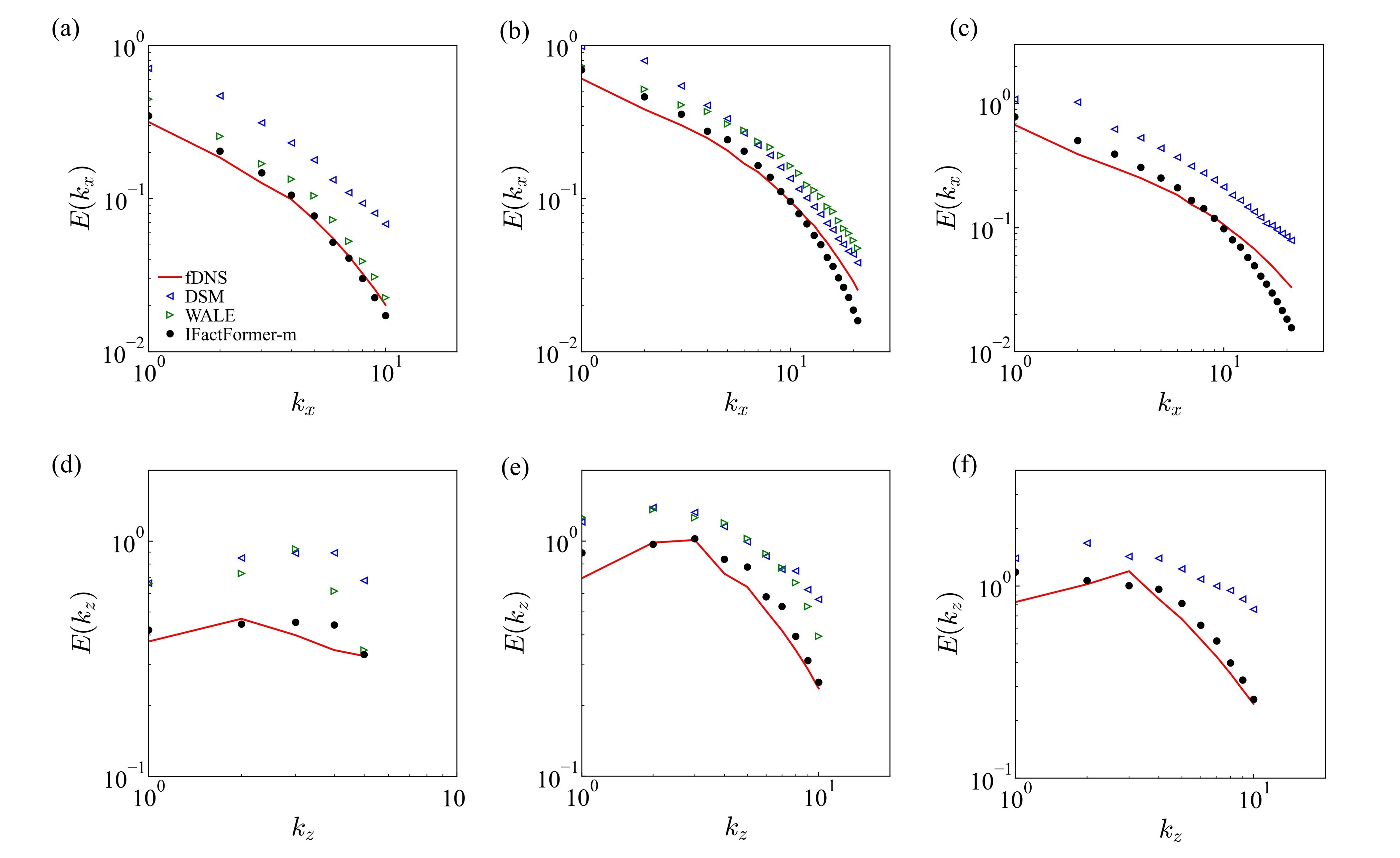}
\caption{The energy spectrum at various Reynolds numbers: (a)-(c) streamwise spectrum at $\text{Re}_{\tau}\approx 180,395,590$;  (d)-(f) spanwise spectrum at $\text{Re}_{\tau}\approx 180,395,590$.}
\label{fig:Ek}
\end{figure*}

\begin{figure*}[!t]
\centering
\includegraphics[scale=0.95]{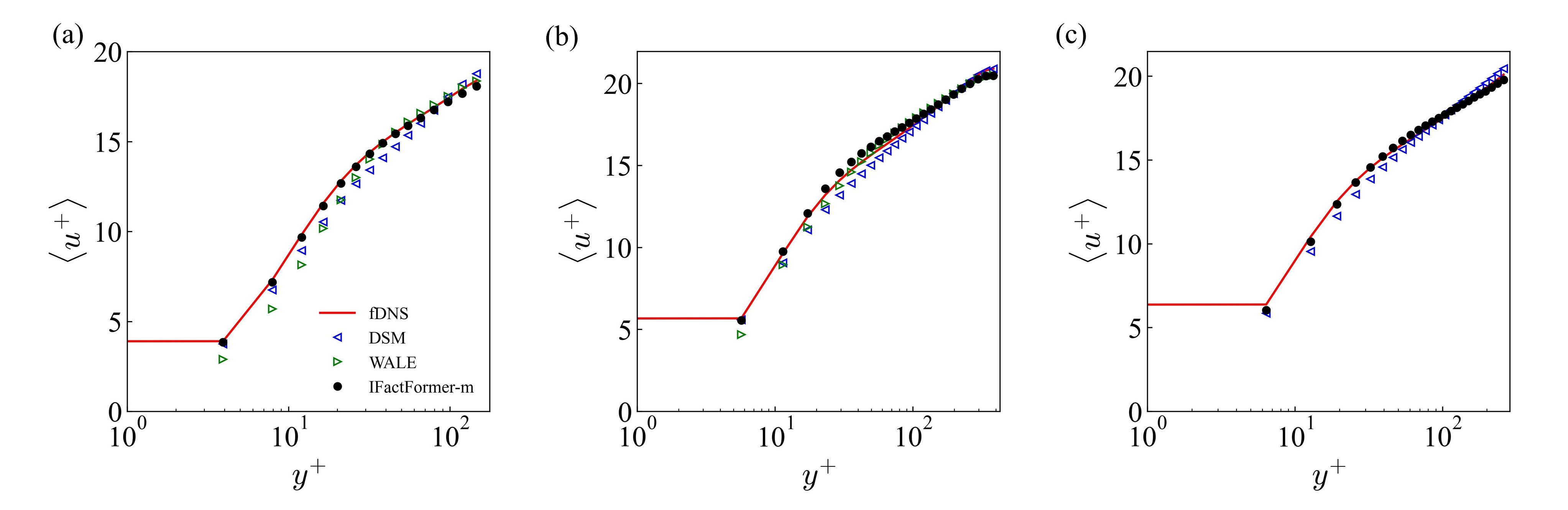}
\caption{The mean streamwise velocity at various Reynolds numbers: (a) $\text{Re}_{\tau}\approx 180$; (b) $\text{Re}_{\tau}\approx 395$; (c) $\text{Re}_{\tau}\approx 590$.}
\label{fig:u+}
\end{figure*}

\begin{figure*}[!t]
\centering
\includegraphics[scale=1]{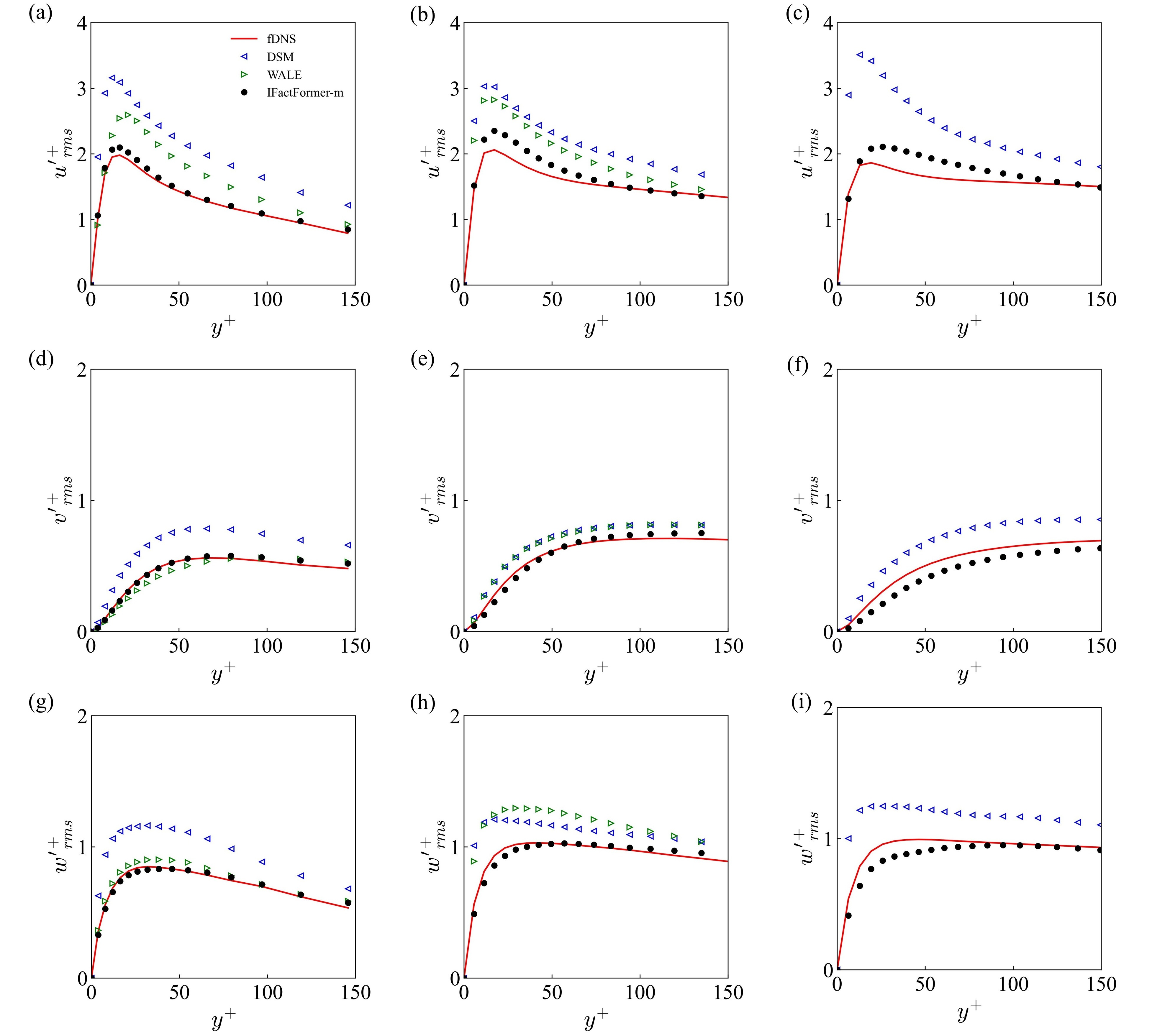}
\caption{The rms fluctuating velocities at various Reynolds numbers: (a)-(c) rms fluctuation of streamwise velocity at $\text{Re}_{\tau}\approx 180, 395, 590$; (d)-(f) rms fluctuation of wall-normal velocity at $\text{Re}_{\tau}\approx 180, 395, 590$; (g)-(i) rms fluctuation of spanwise velocity at $\text{Re}_{\tau}\approx 180, 395, 590$.}
\label{fig:flat}
\end{figure*}

Among the four ML models mentioned above, only the IFactFormer-m model achieves stable long-term predictions. Therefore, subsequent comparisons are made only between the traditional LES method and the predicted results of IFactFormer-m. Due to the divergent behavior of the WALE model in simulating channel turbulence at $\text{Re}_{\tau}\approx 590$ on the grid used in this study, the DSM model is the only one considered as the representative LES model in this case. All comparative results presented below are time-averaged statistics obtained by averaging 400 time steps.

\Cref{fig:Ek} compares the streamwise and spanwise energy spectra of the IFactFormer-m model and the traditional LES model at various Reynolds numbers. At different Reynolds numbers, the energy spectra predicted by the IFactFormer-m model are closer to the fDNS spectra than those calculated by the DSM and WALE traditional LES models. We observe that as the Reynolds number increases, the high-frequency portion of the streamwise energy spectrum predicted by the IFactFormer-m model is relatively smaller than that of the fDNS, indicating that as time steps progress, the error in the IFactFormer model accumulates at the small scales. However, this does not severely impact the prediction of IFactFormer-m of the large scales. For the energy spectrum in the non-dominant direction, while IFactFormer-m outperforms the DSM and WALE models, there is still significant room for improvement. An effective approach could be to use a diffusion model to correct the model's predictions, thereby reducing the errors in the energy spectrum \cite{20}.

\Cref{fig:u+} presents the predicted mean streamwise velocity by the different models at various Reynolds numbers. Both DSM and WALE provide relatively accurate predictions of the mean streamwise velocity, with slight errors in the near-wall region. The IFactFormer-m model, however, can accurately predict the mean streamwise velocity at all locations, almost perfectly overlapping with the fDNS curve.

The root-mean-square (rms) fluctuating velocities are crucial quantities in turbulence characterization, and can be used to measure the intensity or energy of turbulence. \Cref{fig:flat} shows the predicted rms fluctuating velocities in the streamwise, wall-normal, and spanwise directions by different models at various Reynolds numbers. The IFactFormer model accurately predicts the rms velocity fluctuations in all three directions at $\text{Re}_{\tau}\approx 180$, but shows some deviation at $\text{Re}_{\tau}\approx 395$ and $590$. However, the traditional DSM and WALE methods already exhibit significant errors, even at lower Reynolds numbers. This clearly demonstrates that the IFactFormer-m model is capable of predicting turbulence with more realistic intensity at very coarse grids compared to traditional LES methods. 

\begin{figure*}[!t]
\centering
\includegraphics[scale=0.95]{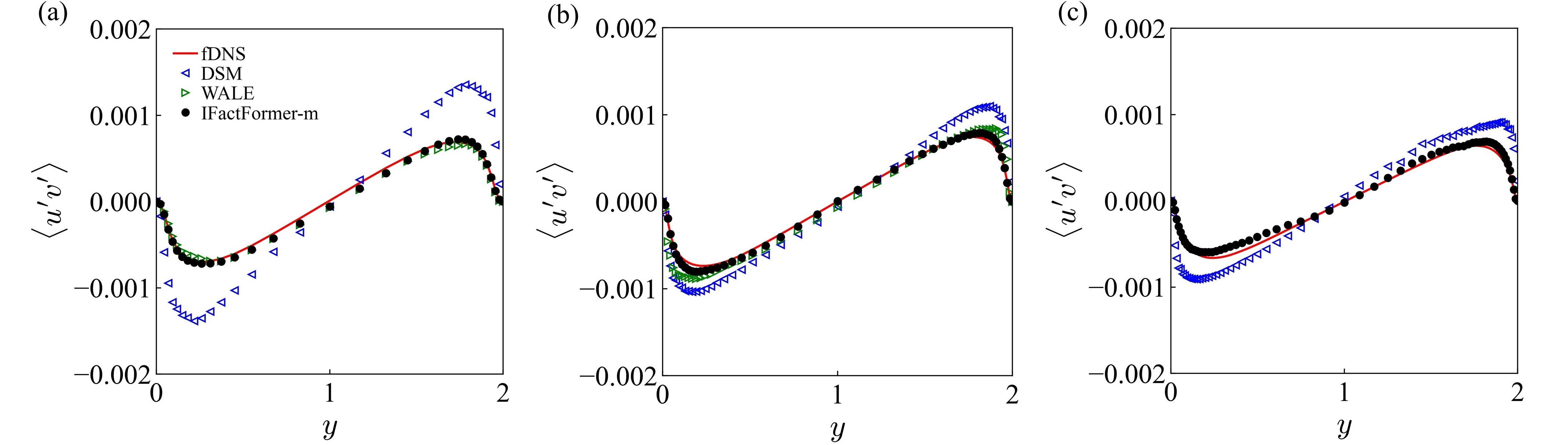}
\caption{The variation of the Reynolds shear stress $\left \langle u' v' \right \rangle $ at various Reynolds numbers: (a) $\text{Re}_{\tau}\approx 180$; (b) $\text{Re}_{\tau}\approx 395$; (c) $\text{Re}_{\tau}\approx 590$.}
\label{fig:Re_stress}
\end{figure*}

\Cref{fig:Re_stress} shows the predicted Reynolds shear stress $\left \langle u' v' \right \rangle $ from IFactFormer-m, DSM and WALE. The Reynolds shear stress $\left \langle u' v' \right \rangle $ exhibits an antisymmetric distribution on either side of the plane $y = 1$. Both IFactFormer-m and WALE accurately predict the location and intensity of the maximum Reynolds shear stress, while the DSM model exhibits notable discrepancies.

\begin{table*}[!t]
\footnotesize
\caption{Computational costs of different models on turbulent channel flow for 80000 DNS time steps (400 ML time steps).}
\tabcolsep 20pt
\begin{tabular*}{\textwidth}{lcccccc}
\toprule
$\text{Re}_{\tau}$ & DSM & WALE & FNO & IFNO & IFactFormer-o & IFactFormer-m \\ 
\hline
180 & 779.2 s & 416.8 s & 17.9 s & 16.6 s & 11.1 s & 18.9 s \\
395 & 2297.6 s & 1170.4 s & 67.0 s & 66.8 s & 23.2 s & 39.4 s \\
590 & 2522.4 s & N/A & 78.8 s & 80.1 s & 31.6 s & 53.4 s \\
\bottomrule
\label{tab:comput_cost}
\end{tabular*}
\end{table*}

\Cref{tab:comput_cost} presents a comparison of the computational costs required to predict 80000 DNS time steps using two LES methods and four ML models. The DSM and WALE models are run on 16, 32, and 64 cores for the three Reynolds numbers $\text{Re}_{\tau} \approx$ $180$, $395$, and $590$, using Intel(R) Xeon(R) Gold 6148 CPUs @ 2.40 Ghz. The four ML models are all run on a single NVIDIA V100 GPU and a CPU configuration of Intel(R) Xeon(R) Gold 6240 CPU @2.60 GHz for inference. The ML models are at least ten times faster in terms of prediction speed compared to traditional LES methods, suggesting that the IFactFormer-m model has the potential to replace traditional LES methods for more accurate and efficient predictions.

\section{Discussion}\label{sec:5}

\begin{figure*}[!t]
\centering
\includegraphics[scale=0.6]{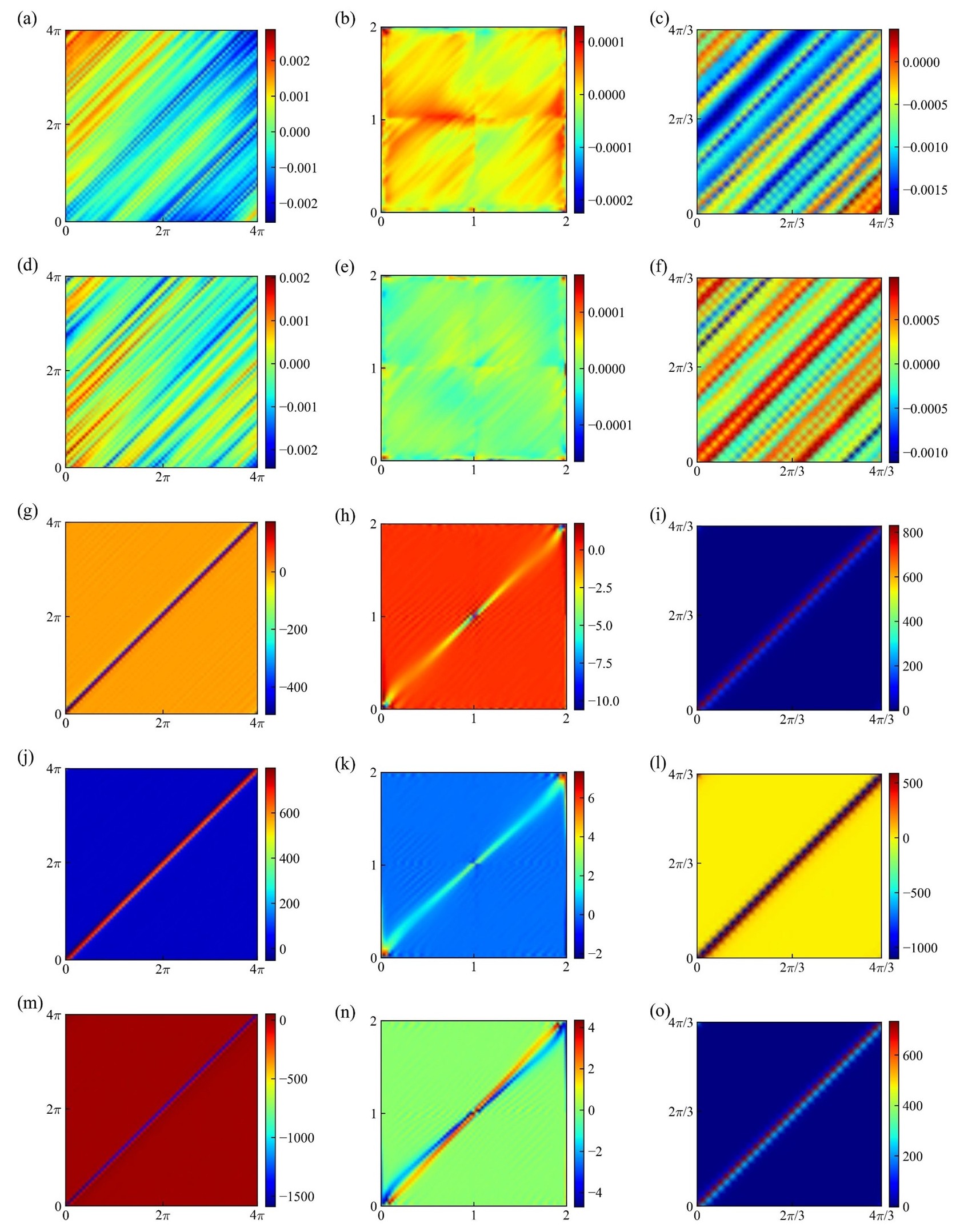}
\caption{Visualization of attention kernel along each axis of IFactFormer-o at $\text{Re}_{\tau}\approx 590$. First column: kernel along $x$ axis (Layer 5, Head 1-5); Middle column: kernel along $y$ axis (Layer 5, Head 1-5); Last column: kernel along $z$ axis (Layer 5, Head 1-5).}
\label{fig:kernel_of_original}
\end{figure*}

\begin{figure*}[!t]
\centering
\includegraphics[scale=0.6]{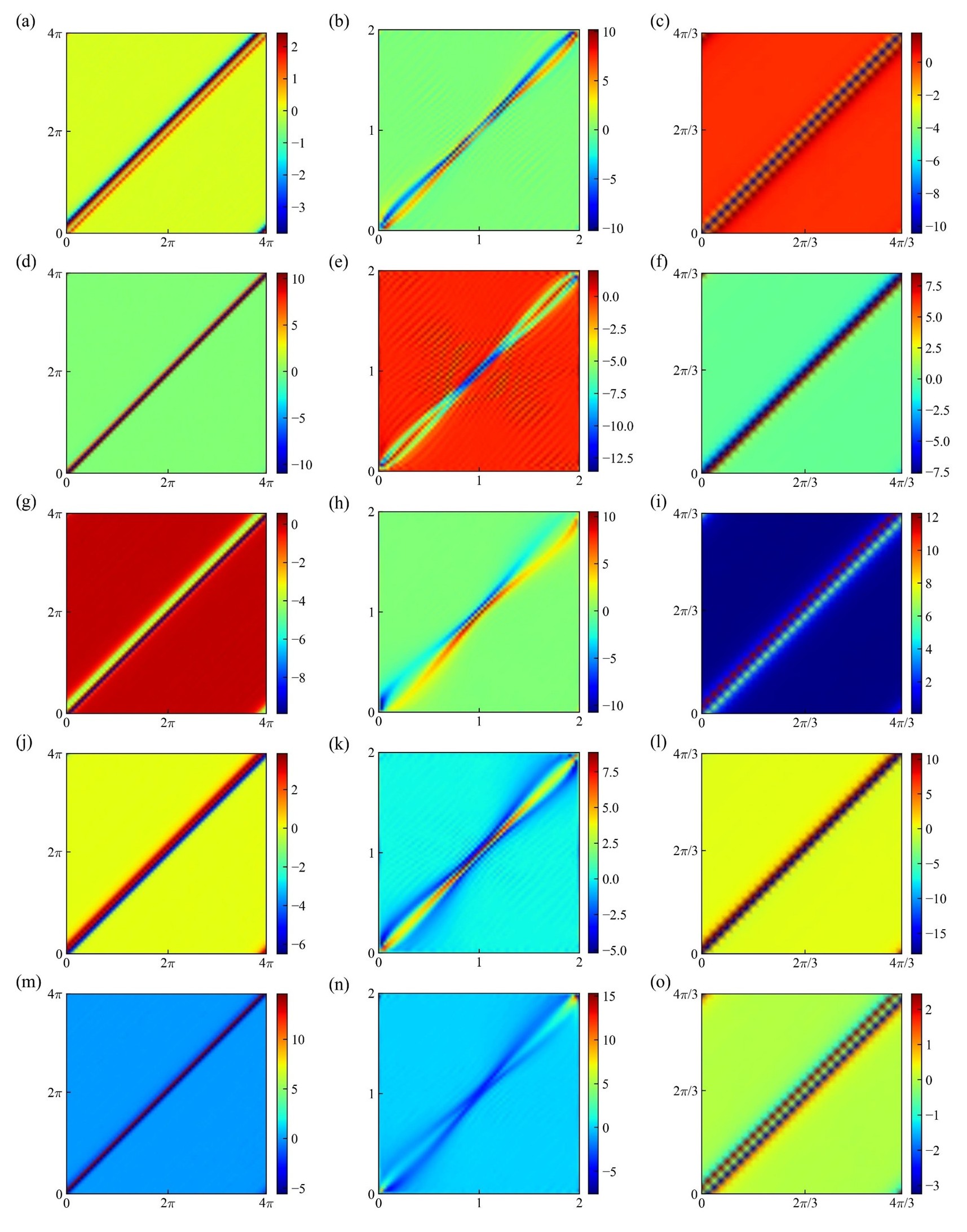}
\caption{Visualization of attention kernel along each axis of IFactFormer-m at $\text{Re}_{\tau}\approx 590$. First column: kernel along $x$ axis (Layer 5, Head 1-5); Middle column: kernel along $y$ axis (Layer 5, Head 1-5); Last column: kernel along $z$ axis (Layer 5, Head 1-5).}
\label{fig:kernel_of_modified}
\end{figure*}

To provide a more detailed comparison and analysis of the chained factorized attention and parallel factorized attention, we visualize the kernel functions along each axis computed in the last layer of the IFactFormer-o and IFactFormer-m models at $\text{Re}_{\tau}\approx 590$. in \Cref{fig:kernel_of_original} and \Cref{fig:kernel_of_modified}.

As illustrated in \Cref{fig:kernel_of_original}(a)-(f), two out of the five attention heads in IFactFormer-o exhibit a ``failure'' phenomenon, where the kernel functions along each axis are nearly zero. This means that these parameters not only fail to extract any meaningful information but also introduce challenges such as slow convergence and increased difficulty in optimization. For the remaining three attention heads, we observe that the kernel functions along the $x$ and $z$ axis exhibit significant numerical jumps in magnitude. This suggests that the IFactFormer-o model learned a ``rigid'' system, which introduces significant numerical instability during time step advancement.

As shown in \cref{fig:kernel_of_modified}, the kernel functions of IFactFormer-m exhibit a high degree of regularity. In the streamwise(first column in \cref{fig:kernel_of_modified}) and spanwise(last column in \cref{fig:kernel_of_modified}) directions, considering the periodic boundary conditions in both, the kernel function is largely independent of the absolute position of the fluid and depends only on the local relative position.  In the wall-normal direction(middle column in \cref{fig:kernel_of_modified}), however, the kernel function is related to the distance of the fluid from the wall and exhibits antisymmetry about the $y=1$ plane. Additionally, from the values of the kernel functions, it can be inferred that IFactFormer-m has learned a ``robust'' system, which is a crucial factor enabling its ability to achieve long-term stable predictions.

These phenomena provide a clear explanation for why IFactFormer-m outperforms IFactFormer-o in terms of convergence speed, short-term prediction accuracy, and long-term prediction stability. The chained factorized attention can achieve satisfactory results in relatively simple isotropic problems. However, when confronted with more complex issues, the flawed structural design increases the difficulty of model representation, preventing the model from learning an appropriate kernel function and consequently leading to a series of problems. In contrast, the parallel factorized attention addresses this issue through minor adjustments.

\section{Conclusions}\label{sec:6}
In this paper, we propose a modified implicit factorized transformer model (IFactFormer-m), which significantly enhances model performance by modifying the original chained factorized attention to parallel factorized attention. Compared to FNO, IFNO, and the original IFactFormer (IFactFormer-o), the IFactFormer-m model achieves more accurate short-term predictions of the flow fields and more stable long-term predictions of statistical quantities in turbulent channel flows at various Reynolds numbers. We further compare the IFactFormer-m model with traditional LES methods (DSM and WALE), using a range of time-averaged statistical quantities, including the energy spectrum, mean streamwise velocity, rms fluctuating velocities, and shear Reynolds stress. The results demonstrate that the trained IFactFormer-m model is capable of rapidly achieving accurate long-term predictions of statistical quantities, highlighting the potential of ML methods as a substitute for traditional LES approaches. Moreover, we analyze the attention kernels of IFactFormer-m and IFactFormer-o, explaining why IFactFormer-m achieves fast convergence and stable long-term predictions.

Current ML models also face a number of challenges, including but not limited to: 

First, one issue is that the IFactFormer-m model currently needs a large amount of fDNS data for training. Physics-informed neural operator (PINO) methods can add physical information into the network \cite{35}. These methods \cite{35,36,37,38,39,40,41,42} have the potential to reduce the model's reliance on high-precision data and are an important direction for future research. Second, another issue is that the predictions of IFactFormer-m on coarse grids cannot support subsequent DNS calculations, as coarse grids fail to meet the Courant-Friedrichs-Lewy (CFL) condition required for DNS. A promising direction to address this limitation is to combine IFactFormer-m with flow field super-resolution models \cite{43,44}. Moreover, generalization is a big challenge. Currently, IFactFormer-m can only predict at a given Reynolds number, geometry, and time step. Developing a machine learning model that can extend to different Reynolds numbers and geometries and be applied to different time steps in the future will be very meaningful. Methods for continuous-time modeling, such as the neural dynamical operator \cite{45}, are expected to enable predictions at different time steps. In addition, designing an effective encoder for Reynolds number and geometry is expected to improve the ML model’s ability to generalize to different Reynolds numbers and geometries.

\Acknowledgements{This work was supported by the National Natural Science Foundation of China (NSFC Grant Nos. 12172161, 12302283, 92052301, and 12161141017), by NSFC Basic Science Center Program (grant no. 11988102), by the Shenzhen Science and Technology Program (Grant No. KQTD20180411143441009), and by Department of Science and Technology of Guangdong Province (Grant No. 2019B21203001, No. 2020B1212030001, and No. 2023B1212060001). This work was also supported by Center for Computational Science and Engineering of Southern University of Science and Technology, and by National Center for Applied Mathematics Shenzhen (NCAMS).}

\InterestConflict{The authors declare that they have no conflict of interest.}


\end{multicols}
\end{document}